\documentclass{article}  
\usepackage{jamaica04}
\usepackage{graphicx}
\usepackage{amsmath}

\frompage{000} \topage{000}                                              

\def\rightmark{3D Imaging in Heavy-Ion Reactions}
\def\leftmark{D.A. Brown et al.}

\newcommand{\dn}[2]{d^{#1}{#2}\,}
\newcommand{\etal}{{\em et al.}}
\newcommand{\rmsubscript}[2]{{#1}_{\textrm{#2}}}

\newcommand{\Rinv}{\rmsubscript{R}{inv}}
\newcommand{\Rs}{\rmsubscript{R}{s}}
\newcommand{\Ro}{\rmsubscript{R}{o}}
\newcommand{\Rl}{\rmsubscript{R}{l}}

\newcommand{\qs}{\rmsubscript{q}{s}}
\newcommand{\qo}{\rmsubscript{q}{o}}
\newcommand{\ql}{\rmsubscript{q}{l}}
 
\title{Three-Dimensional Imaging Analysis of \\ Two-particle Correlations
in Heavy-Ion Reactions} 
\authors{
{D.A. Brown$^1$, P. Danielewicz$^2$, M. Heffner$^1$, R. Soltz$^1$ %
}\\[2.812mm]
{\normalsize
\hspace*{-8pt}$^1$ Lawrence Livermore National Laboratory, Livermore CA, USA\\[0.2ex] 
\hspace*{-8pt}$^2$ Michigan State University, East Lansing MI, USA
}}
 
\abstract{We report an extension of the source imaging method for imaging full three-dimensional sources from three-dimensional like-pair correlations.  Our technique consists of expanding the correlation data and the underlying source function in spherical harmonics and inverting the resulting system of one-dimensional integral equations.  With this method of attack, we can image the source function quickly, even with the extremely large data sets common in three-dimensional analyses.  We apply our method to the recently measured E859 un-Coulomb corrected data.}

\keyword{intensity interferometry, two particle correlations, relativistic heavy-ion collisions, imaging emitting sources} 
\PACS{25.75.-q,25.75.Gz}
 
\begin{document}
 
\maketitle
\setcounter{page}{1}

\section{Introduction}\label{intro}

A single nucleus--nucleus collision at RHIC can produce tens of thousands of particles.  Basic quantum mechanics tells us that {\em all} of these particles are entangled; they are all part of the same final state wavefunction!  The entanglement is a result of conservation laws, particle statistics and inter-particle interactions.  
Intensity interferometry uses pairs entangled by particle statistics and inter-particle interactions to measure the space-time extent of a nucleus-nucleus collision, i.e. the source.  This technique starts with the construction of the correlation function: 
\begin{equation}
    C(\vec{q})=
    \left.
        \displaystyle\frac{dN_\text{pairs}}{d\vec{p}_1d\vec{p}_2}
    \right/
        \displaystyle\frac{dN_\text{mixed}}{d\vec{p}_1d\vec{p}_2}.
\end{equation}
Here, the numerator is the spectrum of particle pairs from the same event and constitutes the signal as these pairs are entangled.  The denominator is the spectrum of particle pairs from different events, using the same cuts as the signal, and constitutes the background as it is constructed to look like the signal without the entangled pairs.  

For identical charged mesons, the correlations are nearly Gaussian except near the origin where the Coulomb force distorts the correlation.  This effect is usually weak and corrected out leaving a correlation one fits to a Gaussian:
\begin{equation}
    C(\vec{q}) = 1+\lambda\exp\left(
        -\frac{\qs^2}{4\Rs^2}-\frac{\qo^2}{4\Ro^2}-\frac{\ql^2}{4\Rl^2}
    \right),\;\;\text{where $\vec{q}=\frac{1}{2}(\vec{p}_1-\vec{p}_2)$}.
\end{equation}
Here the correlation is tabulated in the Bertsch-Pratt coordinates (namely the sideward, outward and longitudinal directions).  Following uncertainty principal-type arguments, the radii of the correlation are taken as the radii of the source.  Unfortunately, the Coulomb correction has a tendency to distort the extracted radii~\cite{mike}.  Usually the distortion is unimportant, but at RHIC we find that the sideward and outward radii are so close that this distortion can make the difference between radii that seem to indicate imaginary freeze-out durations and radii that do not.  

For identical charged mesons, we would like to forget (or at least correct away) final-state interactions.  For other pairs such as proton pairs or proton-$\lambda$ pairs, final-state interactions dominate the signal~\cite{plam}.  The Koonin-Pratt equation provides more general framework~\cite{kp}:
\begin{equation}
     C({\vec{q}}) = \int \dn{3}{r} |\Phi_{\vec{q}}({\vec{r}})|^2 S(\vec{r}).
    \label{eqn:kp}
\end{equation}
Here $S(r)$ is the source function and it gives the probability of producing a pair a distance $\vec{r}$ apart in the pair CM frame.  Also, here $\Phi_{\vec{q}}^{(-)}(\vec{r})$ is the pair's final state wavefunction which we can compute if we know their interaction.  In this paper, we will refer to the square wavefunction as the kernel, $K(\vec{q},\vec{r}) = |\Phi_{\vec{q}}^{(-)}(\vec{r})|^2$.  

Since we can no longer read off the radii to get the source size, we must invert Eq.~\eqref{eqn:kp} to get $S(\vec{r})$ directly.  We will do this with an extension of the one-dimensional source imaging technique introduced in Refs.~\cite{imag}.  Source imaging allows us to cleanly separate the effects due to final-state interactions and symmetrization from effects due to the source function itself.  Because we make no assumptions about the functional form of the source function, the imaging can reveal non-Gaussian sources.  In the remainder of this paper, we will outline the three-dimensional imaging process.  We first state our test problem, then detail the imaging process and finally show the imaging results on the test problem.  With this experience in hand, we analyze data from the AGS-E859 experiment.


\section{Setting Up the Problem}\label{imaging}  

\begin{figure}[htb]
    \begin{minipage}[t]{2.075in}
        \includegraphics[width=6cm]{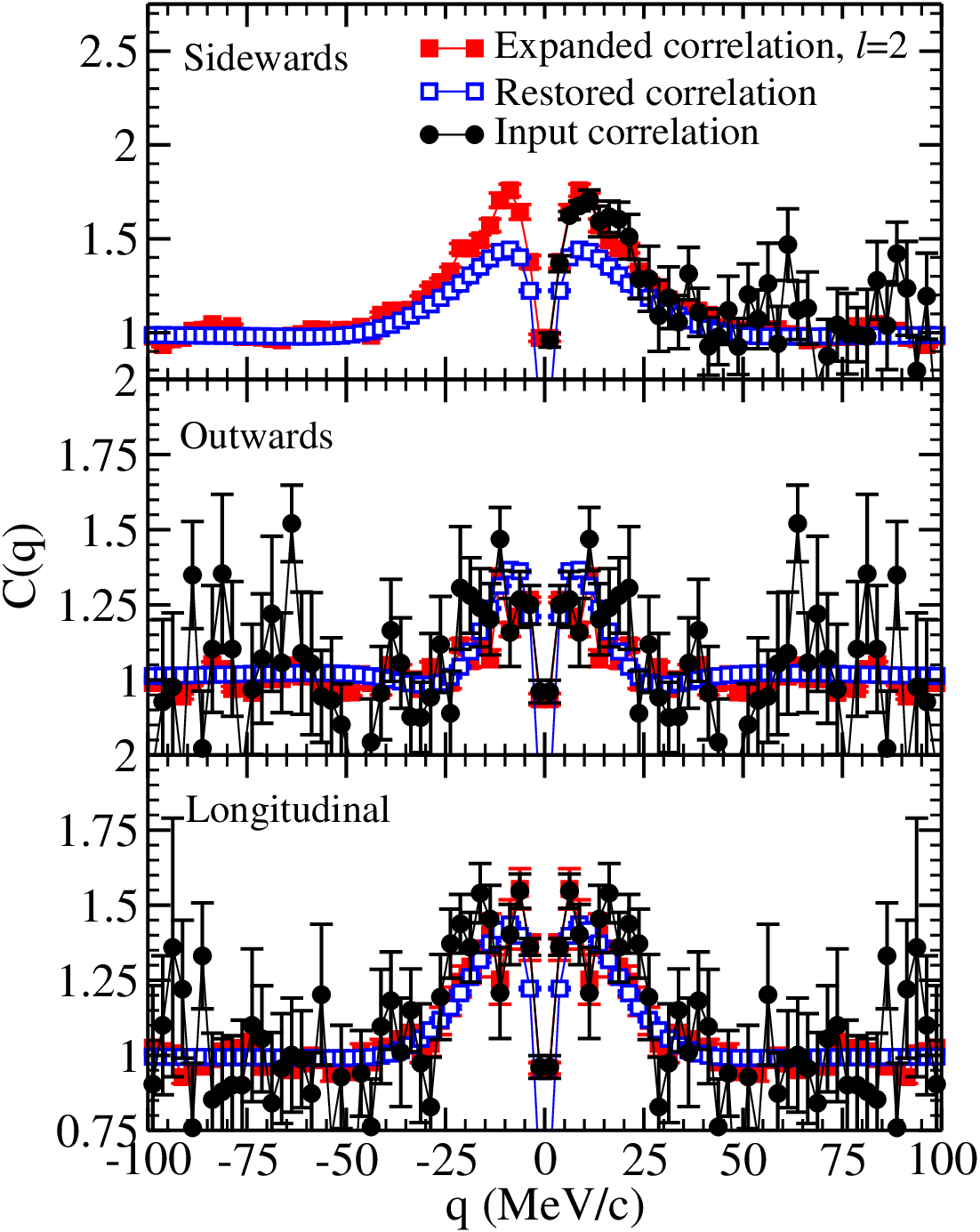}
    \end{minipage}\ \hspace{0.35in}
    \begin{minipage}[t]{2.075in}
        \includegraphics[width=6cm]{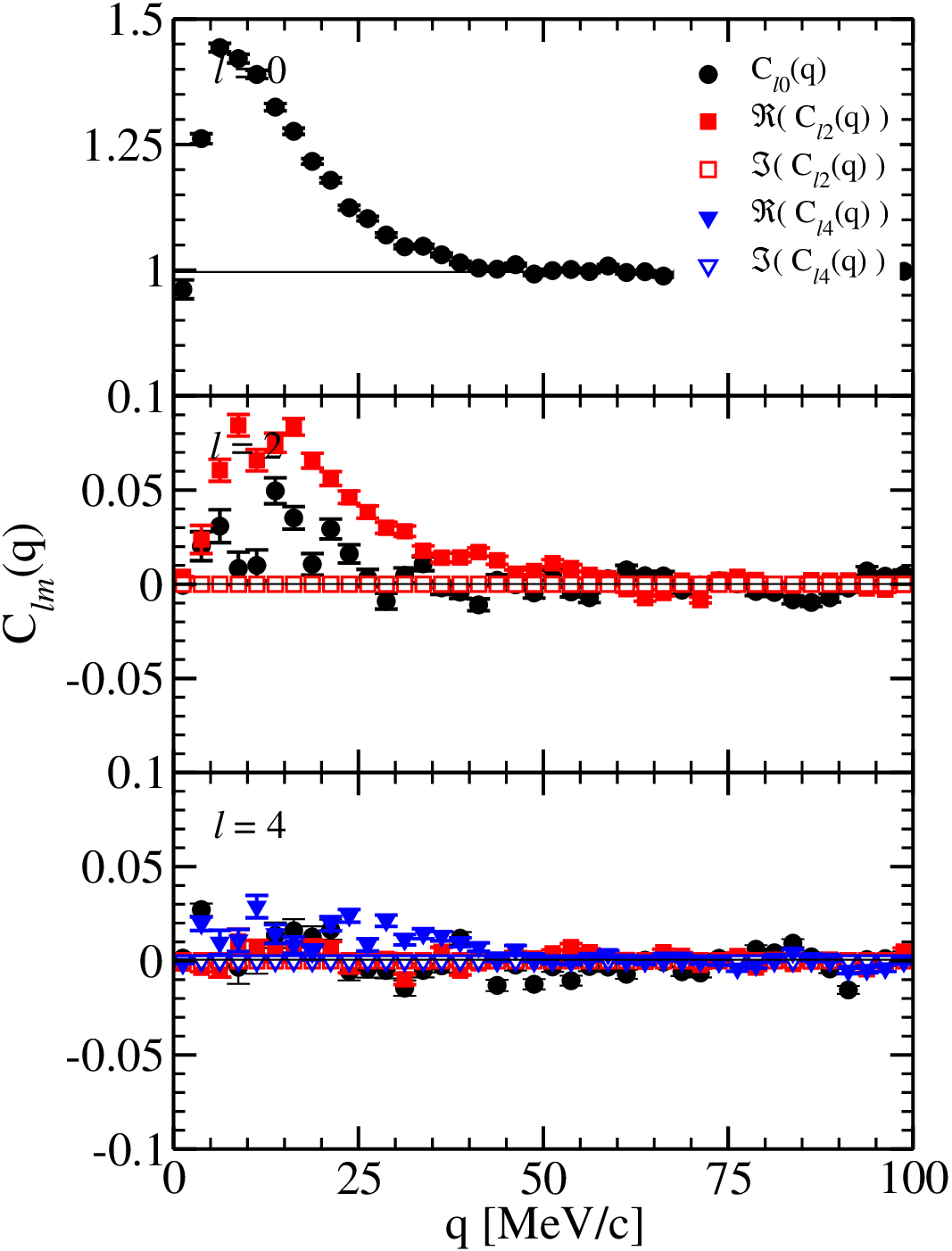}
    \end{minipage}
    \caption[]{Left panel: 2.5 MeV slices of the full three-dimensional test correlation.  The different curves are the test correlation (black), the correlation expanded in Spherical Harmonics (red), and the imaged, then unimaged, correlation (blue).  Right panel: terms in the Spherical Harmonic expansion of the test correlation.}
    \label{fig1}
\end{figure}

We begin with an emission function that has finite lifetime ($\tau=20$~fm/c\footnote{We choose $\tau$ to be in the ballpark of the $\omega$ meson lifetime (23 fm/c).}), finite temperature ($T=50$~MeV), and a Gaussian spatial profile ($R_x,R_y,R_z=4$~fm):
\begin{equation}
   D(\vec{r},t,\vec{p})\propto
   \exp\left(-t/\tau\right)
   \exp \left(-\frac{{\vec{p}\;}^2}{2m_\pi T}\right)
   \exp\left(-\frac{x^2}{R^2_x}-\frac{y^2}{R^2_y}-\frac{z^2}{R^2_z}\right).
\end{equation}
On the face of it, this source would seem to produce an unchallenging spherically symmetric source function.  However since we must boost from the emission function's frame (i.e. the lab frame) to the pair CM frame, the finite lifetime will translate into a non-Gaussian source function through:
\begin{equation}
   S(\vec{r'})=\int \dn{}{r_0'} \int \dn{4}{R} D(R+r/2,\vec{p}_1) D(R-r/2,\vec{p}_2).
\end{equation}
With the emission function specified, we combine it with the $\pi^+\pi^+$ kernel to get a correlation using Scott Pratt's CRAB code.  This kernel includes both the correct pair symmetrization and the Coulomb force between the pair.  On the left panel of Fig.~\ref{fig1}, we show 2.5 MeV wide slices of the correlation along the sidewards, outwards and longitudinal axis.

To perform the imaging analysis, we convert the three-dimensional Koonin-Pratt Equation in Eq.~\eqref{eqn:kp} to a series of one-dimensional equations:
\begin{equation}
	C_{\ell m}(q) - \delta_{\ell 0} = 4\pi\int_0^\infty \dn{}{r} r^2 K_\ell(q,r) S_{\ell m}(r).
    \label{eqn:kp1d}
\end{equation}
To do this, we simply expand the source function and correlation function in Spherical Harmonics, expand the kernel in Legendre polynomials, and then apply the Spherical Harmonic Addition Theorem: 
\begin{equation}
	C(\vec{q})=\sqrt{4\pi}\sum_{\ell=0}^{\ell_{max}} \sum_{m=-\ell}^{\ell}C_{\ell m}(q) Y_{\ell m}(\hat{q}),\;\;\;\;
    S(\vec{r})=\sqrt{4\pi}\sum_{\ell=0}^{\ell_{max}} \sum_{m=-\ell}^{\ell}S_{\ell m}(r) Y_{\ell m}(\hat{r}),
\end{equation}
\begin{equation}
	\text{and}\;\; K(\vec{q},\vec{r}) = \sum_{\ell=0}^{\ell_{max}}(2\ell+1)K_{\ell}(q,r) P_{\ell}(\hat{q}\cdot\hat{r}).
\end{equation}
The right panel in Fig~\ref{fig1} shows the first few terms in the expansion of the test problem correlation.

Having done this, it is natural to ask what these terms mean~\cite{pratt_interp}.  The $\ell = 0$ term is the angle averaged correlation and gives us access to $\Rinv$.  The $\ell = 1$ terms give us access to Lednicky offset, in other words which particle was emitted first.  These terms are absent from like pair correlations due to symmetry.  The $\ell = 2$ terms give us shape information and access to $\Ro$, $\Rs$, $\Rl$: $C_{20} \sim R_{L}$ and $\alpha_{\pm}C_{00}-(C_{20}\pm C_{22}) \sim R_S ,R_O$ (here $\alpha_{\pm}$ is a constant of order one).  Finally, $\ell = 3$ terms give us access to ``boomerang'' or triaxial deformation and are absent from like pair correlations due to symmetry.   Terms with $\ell \ge 4$ probably will be found in future datasets, but are not present in either our test problem or in the real data we have analyzed.

In order to image the test correlation, we must complete converting the one-dimensional equations in Eq.~\eqref{eqn:kp1d} to matrix equations.  We do this by representing the radial dependence of the source terms in Basis splines:
\begin{equation}
   S_{\ell m}(r)=\sum_{j=1}^{N_c}S_{j\ell m} B_j(r).
\end{equation}
Basis splines are piece-wise continuous polynomials and the polynomials that make up the splines are patched together at the spline knots.  Setting the spline knots is both crucial for representing source well and difficult to do.  Our method for doing this is beyond the scope of this paper and will be discussed in~\cite{inprog}.

Inverting the series of matrix equations in Eq.~\eqref{eqn:kp1d} is an ill-posed problem: the kernels are not square and possibly singular, the data is noisy, and error propagation is crucial.  The practical solution to linear inverse problem, is to find the vector of source coefficients that minimize the $\chi^2$:
\begin{equation}
   \chi^2=(K\cdot\vec{S}-\vec{C})^T\cdot
      ({\Delta^2 C})^{-1}\cdot(K\cdot\vec{S}-{\vec{C}}).
\end{equation}
The source that does this is $\vec{S}={\Delta^2 S}\cdot K^T\cdot(\Delta^2 C)^{-1}\cdot{\vec{C}}$ with the source covariance matrix ${\Delta^2 S} = (K^T\cdot (\Delta^2 C)^{-1}\cdot K)^{-1}$.


\section{Data Analysis}\label{analysis}

\begin{figure}[htb]
    \begin{center}
    \includegraphics[width=7cm]{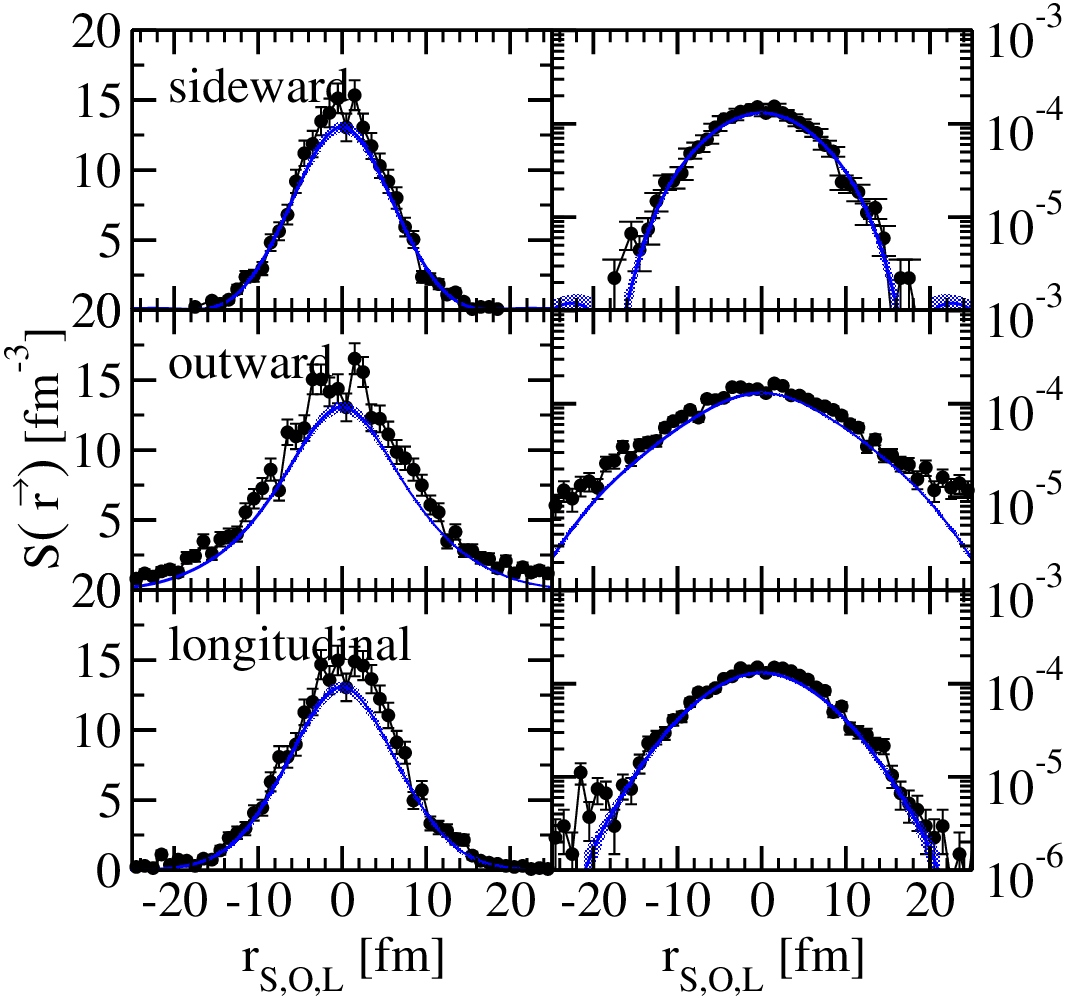}
    \end{center}
    \caption[]{Comparison of the imaged source function (blue) and the input source function (black).  The left set of panels are on a linear scale and the right set of panels are on a log scale.}
    \label{fig2}
\end{figure}

Now we present the results of the imaging analysis on the test problem.  In Fig.~\ref{fig2} we show slices through the imaged source function and the input source function.  Clearly we were able to match the height, width and integral of the source function.  We even were able to begin to resolve the non-Gaussian halo in the longitudinal and outwards direction caused by the combination of the finite lifetime of the underlying emission function and the lab to pair CM boost.  In addition, we can simply read off the equivalent Gaussian radii from the source plots and find: $\Rs = 4.15 \pm 0.08$ fm, $\Ro = 5.17 \pm 0.12$ fm, and $\Rl = 4.53 \pm 0.09$ fm.  We can also read off the height to get $S(r=0)= 13 \pm 3\times 10^{-5} \text{fm}^{-3}$.  If we integrate the source out to the edge of the imaged region, we find the fraction of pairs that we can resolve is $\lambda= 0.566 \pm 0.025$.  Clearly, we can extract all of the 
important physics directly from these plots.

\begin{figure}[htb]
    \begin{minipage}[c]{2.075in}
        \includegraphics[width=5.5cm]{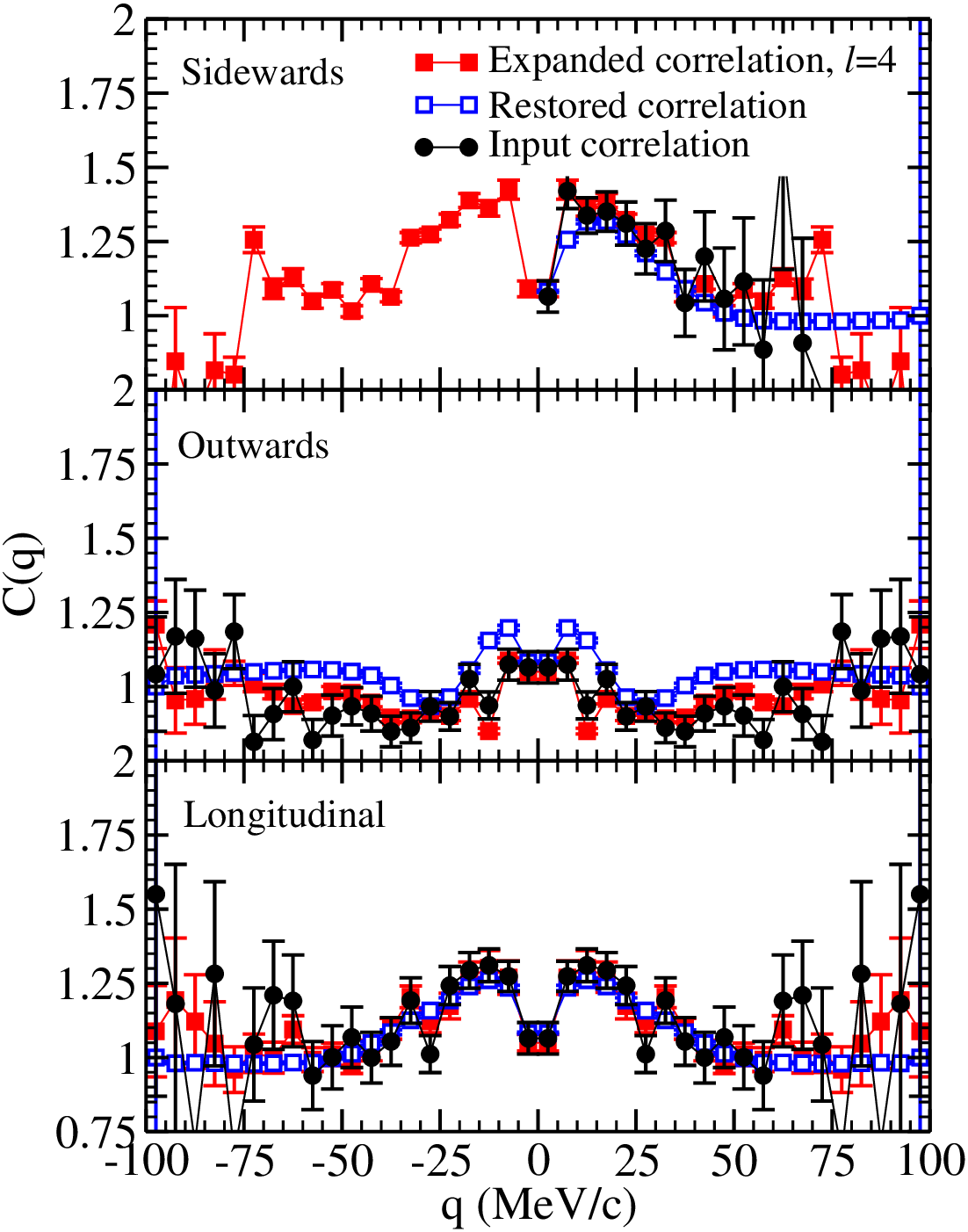}
    \end{minipage}\ \hspace{0.35in}
    \begin{minipage}[c]{2.075in}
        \includegraphics[width=6cm]{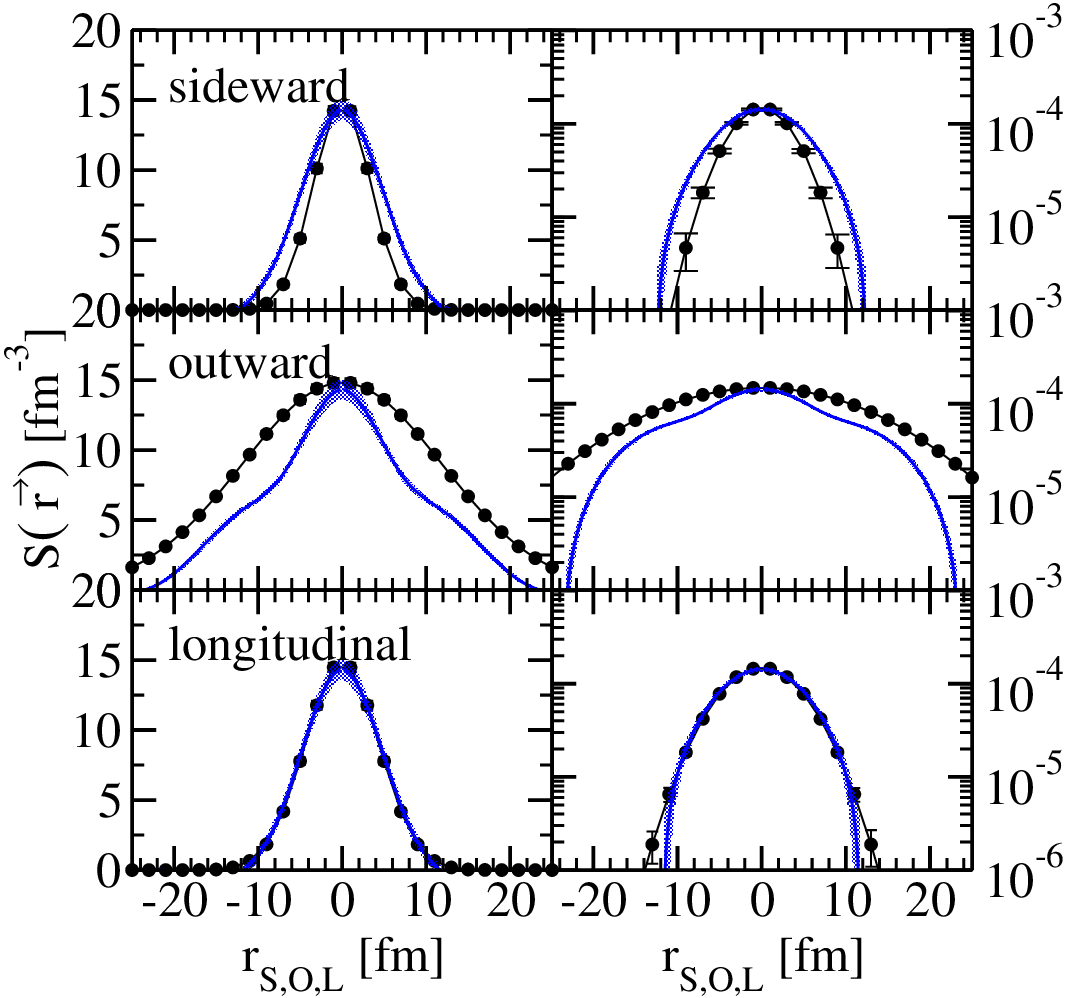}
    \end{minipage}
    \caption[]{Left panel: 5 MeV wide slices of the correlations showing the input correlation (black), input correlation expanded in Spherical Harmonics (red), and imaged then unimaged correlation (blue).  Right panel: slices through the imaged source (blue band) and the best-fit Gaussian from E859 (black).  The left set of sub-panels are the source on a linear scale and the right set of sub-panels are the source on a log scale.}
    \label{fig3}
\end{figure}

Finally we apply the imaging to the data from AGS-E859~\cite{e859}.  This experiment was 14.6 AGeV Si+Au and the data was compiled in the pair CM frame.  Fig~\ref{fig3} shows both the correlation slices and the imaged source slices.
Looking at the imaged source, we see significant differences between the imaged source and E859's best-fit Gaussians.  The Gaussian fits and the images mostly agree in the sideward and longitudinal directions however the outward direction seems to have a two-component structure.   If this structure is real, the simplest interpretation is that the underlying emission function has an associated lifetime in the longitudinally co-moving frame ({\em not} the lab frame).
Unfortunately, as we see in the correlation plots, there may be problems with the data that foil the imaging.  There is a large band of correlation values substantially below one from $\qo=25-75$~MeV/c along the outwards axis.  This is a large unphysical anti-correlation that appears to be an experimental artifact~\cite{ron_private} to which the Gaussian fits are insensitive.


\section{Conclusions}\label{concl}
We have a lot of work to do on this project before it is ready.  In addition to looking to other datasets, such as the recent PHENIX Run2 data, we are finishing the merger of the imaging code with Mike Heffner's {\tt CorAL} code and Scott Pratt's kernel code.  In the future we hope to tackle more general test source functions and emission functions and to be able to handle un-like particles.
  
 
\section*{Acknowledgement(s)}
This work was performed under the auspices of the U.S. Department of Energy 
by University of California, Lawrence Livermore National Laboratory under 
Contract W-7405-Eng-48.
 

\vfill\eject

\begin{thebibliography}{99}  
\bibitem{mike}M. Heffner for the PHENIX Collaboration, 
    Quark Matter 2004 Proceedings, Oakland CA, Jan. 11-17 2004; 
    to be published in the Journal of Physics G.
\bibitem{plam}P. Chung, \etal, Phys. Rev. Lett. {\bf 91}, 162301 (2003).
\bibitem{kp}S.E.~Koonin, Phys. Lett. {\bf B70}, 43 (1977); 
    S.~Pratt, T.~Cs{\"o}rg\H{o} and T.~Zim\'{a}nyi, Phys. Rev.~C {\bf 42}, 2646 (1990). 
\bibitem{imag}D.A.~Brown and P.~Danielewicz,  
   Phys.~Lett.~B {\bf 398}, 252 (1997);
   Phys.~Rev.~C {\bf 57}, 2474 (1998);
   Phys. Rev.~C {\bf 64}, 014902 (2001).
\bibitem{crab}CRAB webpage:
   {\tt http://www.nscl.msu.edu/$\tilde{\;\;\;}$pratt/freecodes/crab/home.html}
\bibitem{pratt_interp}S.Pratt, private communication.
\bibitem{inprog}D.A. Brown,  P. Danielewicz, M. Heffner, R. Soltz, in       
    preparation.
\bibitem{e859}L. Ahle, \etal, Phys. Rev. C {\bf 66}, 054906 (2002).
\bibitem{ron_private}R.Soltz, private communication.
\end{thebibliography}
\end{document}